\documentstyle[epsf]{article}
\def\bq{\mbox{\boldmath$q$}}
\def\br{\mbox{\boldmath$r$}}
\def\bsig{\mbox{\boldmath$\sigma$}}
\begin{document}
\begin{center}
{\Large \bf What do mesons have to do with nuclear structure?}
\end{center}
\vspace{0.3cm}

\begin{center}

{Daniel S. Koltun}
\end{center}
\vspace{0.3cm}

{\it Department of Physics and Astronomy, University of Rochester}
\vspace{0.5cm}

{\bf Abstract}
\vspace{0.5cm}
  
The theory of nuclear structure (binding, low energy spectra, transitions, 
etc.) depends on nucleon-nucleon ($NN$) interactions. The meson theory of
$NN$ interactions has predictive power for $NN$ scattering, and partial
success when applied to the theory of nuclear structure.
Is it possible to test this theoretical picture -- by direct experimental
interaction with mesons in nuclei? Some experimental searches for the `pion 
excess' in nuclei have ambiguous results. How sensitive are such experiments 
to the mesonic aspects of nuclear structure? These questions are addressed in 
this talk. More details and references are given in \cite{aa1}.
\vspace{1.0cm}

\section{Meson exchange interactions in nuclei}

The one pion exchange (OPE) potential for momentum transfer $\bq$ is given by
\begin{equation}
V_\pi(q) = -{f^2\over {m_\pi}^2}{(\bsig_1 \cdot \bq)(\bsig_2 \cdot \bq)\over
(q^2 + {m_\pi}^2)}\tau_1 \cdot \tau_2.
\end{equation}
The analogous (spin) potential for exchange of one rho meson can be written
\begin{equation}
V_\rho(q) = -{{f_\rho}^2\over {m_\rho}^2}{(\bsig_1\times \bq)\cdot
(\bsig_2\times \bq)\over(q^2 + {m_\rho}^2)} \tau_1\cdot \tau_2.
\end{equation}
These are both important contributors to the NN tensor force and to nuclear 
binding. For a nucleus $A$, the meson contributions are

\begin{equation}
<V_{\rm meson}>_A = \int {d^3q\over(2\pi)^3}<V_{\rm meson}(q)>_A,
\end{equation}
where, for the pion
\begin{equation}
<V_\pi(q)>_A = -{f^2\over 2m_\pi^2}{S_L^{(2)}(q)\over(q^2 + m_\pi^2)},
\end{equation}
with the longitudinal spin correlation function given by the expectation value
in the nuclear ground state 
\begin{equation}
S_L^{(2)}(q) = {1\over 3Aq^2} \sum^A_{k\neq j} <(\bsig_k\cdot 
\bq)(\bsig_j\cdot \bq) \tau_k\cdot \tau_j 
{\rm exp}[i\bq\cdot(\br_k-\br_j)]>_A.
\end{equation}
Similarly, the rho contribution depends on the transverse spin correlation
function
\begin{equation}
S_T^{(2)}(q) = {1\over 6Aq^2} \sum^A_{k\neq j} <(\bsig_k\times \bq)\cdot
(\bsig_j\times \bq) \tau_k\cdot \tau_j 
{\rm exp}[i\bq\cdot(\br_k-\br_j)]>_A.
\end{equation}

For example, calculations with the Argonne-Urbana 
$NN$ potentials and full correlations \cite{ba} give
\begin{equation}
<V_\pi>_A/A = -30.7\ \rm{MeV}
\end{equation}
for $^{16}$O. This is a major part of the nuclear binding.

\section{Inelastic excitation of nuclei}

Nuclear structure enters inelastic scattering through response functions,
defined as follows:
\begin{equation}
R^a_\alpha (q,\omega) = \sum_f |<f| \Gamma_\alpha^a
(\mbox{\boldmath$q$})|i>|^2 \delta(\omega - E_f + E_i),
\end{equation}
where $i$ and $f$ denote initial and final nuclear states.
In the present context we discuss two:
the longitudinal $R_L^a(q,\omega)$ and transverse $R_T^a(q,\omega)$, which
correspond to exchange (between the projectile and the nuclear target) of 
$0^-$ and $1^-$ quanta, pion-like, or rho-like.
The $\Gamma_\alpha^a(\mbox{\boldmath$q$})$ are given by the following 
single-nucleon operators:

\begin{equation}
\Gamma_L^a(\bq) = \sum_{k=1}^A (\bsig_k \cdot \bq)\tau^a_k \quad 
{\rm exp} (i\bq \cdot \br_k),
\end{equation}
\begin{equation}
\Gamma_T^a(\bq) = \sqrt{1\over 2} \sum^A_{k=1} (\bsig_k \times
\bq) \tau^a_k \quad {\rm exp} (i \bq \cdot \br_k),
\end{equation}

a) Spin transfer to a nuclear target by the $(\vec p,\vec n)$ reaction
can be analyzed, under some assumptions about the reaction, to give
two response functions, $R^+_\alpha(q,\omega)$, with $a=+$.
For a $T=0$ target, these are related to the full isovector response 
functions by $R_\alpha(q,\omega) \equiv \sum_a R_\alpha^a(q,\omega)
= 3R^+_\alpha (q,\omega)$.

b) Inelastic scattering on virtual pions (or other mesons) can be related to
the imaginary part of the forward amplitude
\begin{equation}
{\cal I}m\ {\cal F}(0) = -{1\over 4\pi}\int {d^4q\over (2\pi)^4}P(q) 
{\cal I}m\ {\cal M}(p,q),
\end{equation}
using $p,q$ for the 4-momenta of the projectile and the meson. Here $P(q)$
is an invariant distribution function for the virtual meson, and ${\cal M}$ 
is the invariant scattering amplitude on the meson.
Sullivan gave the result for pions \cite{ac}, which can be written for 
pseudovector coupling to nucleons with form 
factor $F(t)$:
\begin{equation}
P(q) = \pi{f^2\over m_\pi^2}F^2(t) {R_L(q,\omega)\over 
(t + m_\pi^2)^2},
\end{equation}
where $t = q^2 - \omega^2$. Thus the pion contribution is related to the
response $R_L$.

\section{ Pion distribution functions and pion excess}

Pion distributions are connected to nuclear structure
through the response functions.
The pion probability distribution in the nuclear target ($A$) 
is given by: 
\begin{equation}
n_A(q) = <\sum_a a^+_a(\bq) a_a(\bq)>_A.
\end{equation}
More of interest for nuclear structure is the {\em excess} distribution, 
defined by the difference $\delta n_A(q) = n_A(q) - A n_N(q)$,
where $n_N(q)$ is  for a single nucleon. 
This quantity is not directly measured in experiment, but it is connected to
the function $\delta R_L = R_L^{(A)} - A R _L^{(N)}$. In a static 
approximation, and assuming pseudovector coupling of pions to nucleons, 
one finds 
\begin{equation}
\delta n_A(q) = {f^2F^2(q^2)\over 2 \varepsilon^3_q m_\pi^2} 
\int^\infty_0 d\omega \delta R_L(q,\omega) = {3Aq^2f^2F^2(q)\over 2 
\varepsilon^3_q m^2} S_L^{(2)}(q)
\end{equation}
with pion energy $\varepsilon_q = \sqrt{q^2 + m_\pi^2}$.
The last term follows from the sum rules which connect the
response functions to the correlation functions: 
\begin{eqnarray}
S_\alpha(q) & \equiv & {1\over 3Aq^2}\int^\infty_0 d\omega R_\alpha(q,\omega)
\nonumber\\
	    & = & {1\over 3Aq^2} <\sum_a \Gamma_\alpha^{a \dagger}(\bq)
\Gamma_\alpha^a (\bq)>_A = 1 + S_\alpha^{(2)}(q).
\end{eqnarray}
To the same approximation,
$\delta n_A(q) = -{<V_\pi(q)>_A/\varepsilon_q}$.
The integrated excess per nucleon is $\delta n_A/A \simeq 0.03$ 
in the theory which gives (7) \cite{ba,ad}. 

A  quantity which is used in the analysis of DIS experiments is the momentum 
distribution function for pions in the target. This may be defined, using
the scaling variable $y$,  
\begin{equation}
p_A(y) = y\int {d^4q\over (2\pi)^4} P(q) \delta(y - {(q_z - \omega)\over M}),
\end{equation}
with $M$ the nucleon mass and $q_z$ the longitudinal component of $\bq$ .

\section{ What is known about response functions?}

The response of a noninteracting Fermi gas (FG) of nucleons
is given by an inverted parabola in $\omega$\ (for $q \geq 2k_F$), 
whose peak value (when normed to unity) is given by
\begin{equation}
\max {R^{FG}\over3Aq^2} = 
{3M\over 4k_Fq},\ \  {\rm at \ } \omega = {q^2\over 2M},
\end{equation}
and a range $\Delta\omega = 2qk_F/M $.
For the mean-field shell model (SM) the response is qualitatively similar to
that for the FG, governed by the particle-hole spectrum. The peak position 
is usually shifted above the FG value of $\omega = q^2/2M$.

For interacting nuclear systems in general, calculation of the response
functions is difficult. However, the two sum-rule functions $S_\alpha(q),
W_\alpha(q)$ can be obtained as ground state expectation values.
$S_\alpha$ is given by (15); the energy-weighted sum rule is given by
\begin{equation}
W_\alpha(q) = {1\over3Aq^2}\int^\infty_0 d\omega \omega R_\alpha(q,\omega)
= {1\over6Aq^2}<\sum_a[\Gamma_\alpha^{a \dagger}(\bq),[H,\Gamma_\alpha^a(\bq)]
>_A.
\end{equation}
These have been calculated for various nuclear ground states by the 
Argonne-Urbana group, using realistic NN interactions \cite{ad}.

Direct calculation of the response has 
been done for a few cases:

In RPA \cite{bf,aa4}, with $\pi$- and $\rho$-exchange, with strong attraction at
short range (Landau $g' \simeq 0.7$) one gets an enhanced FG or SM peak for
$R_L$, suppressed peak for $R_T$, but no broadening, since the same p-h
spectrum as in the SM dominates the energy dependence.
This reflects the sum rule behavior, $S_L \geq S^{SM} \geq S_T$,
for the implied tensor correlations.

In the Correlated Basis Function method \cite{aa2}, NN correlations are
included in the ground and p-h excited states. The response peaks are shifted
and broadened by the correlations, but the spectrum is still based on the p-h 
assumption. The strengths $S_L$ and $S_T$ are similar to the Argonne-Urbana 
results.

Both above theories simulate the mixing of higher configurations (2p-2h)
by broadening the p-h energies. This adds a high-energy tail to the 
$\omega$ dependence.

A direct calculation of a scalar response function for $^4$He has been done
by an integral transform method \cite{aa3}, but only for central NN 
interactions, which is not suitable for the spin-dependent $R_L, R_T$. 
However, these calculations give large 
strength for $\omega$ well above the SM peak - a high-energy tail.
Additional evidence of strength at high energy is given by \cite{ad}, who 
calculate the Euclidean (Laplace) transforms of both $R_L$ and $R_T$, and for 
$A\ >$ 4.

\section{ Sum rules and models of response functions}

Here I describe a model for the functional dependence of the response 
functions, using information directly from the two sum rules, which are 
more easily calculated than the response itself (See \cite{aa1}).  
The basic idea is to separate the response functions into two components,
the first of which characterizes the noninteracting part, and which gives 
the SM peak. A second distribution includes the effects of correlations,
and provides strength for $\omega$ above the SM peak. We write
\begin{eqnarray}
R_\alpha(q,\omega) & = & R_\alpha^{SM}(q,\omega)+\Delta R_\alpha(q,\omega),\\
S_\alpha(q) & = & S^{SM}(q) + \Delta S_\alpha(q),\\
W_\alpha(q) & = & W^{SM}(q) + \Delta W_\alpha(q) .
\end{eqnarray} 
One can calculate $S_\alpha$ and $W_\alpha$ for interacting nuclei \cite{ad},
and $S^{SM}$ for shell model nuclei. $W^{SM}$ is more model dependent, 
but can be estimated. Then $\Delta S_\alpha$ fixes the integrated strength,
and $\Delta W_\alpha$ the centroid, of the correlation response $\Delta 
R_\alpha$. One needs to assume the functional forms. 
Values of these quantities are given for $^{16}$O in Tables I and II of 
\cite{aa1}.

For $\Delta R_L$ we consider two models:
in the first we take the distribution to be constant, 
e.g., for 0 $< \omega < 2\omega_L(q)$:
\begin{equation}
{\Delta R_L(q,\omega)\over3Aq^2} = {\Delta S_L(q)\over2\omega_L(q)}
\Theta(2\omega_L(q) - \omega).
\end{equation}
This model has a symmetric distribution in $\omega$ about the centroid 
$\omega_L(q) = \Delta W_L(q)/\Delta S_L(q)$. 
The second distribution is not symmetric:
\begin{equation}
{\Delta R_L(q,\omega)\over3Aq^2} = 
{\Delta S_L(q)\over\beta^2}\omega \exp{(-\omega/\beta)}.
\end{equation}
with $2\beta = \omega_L
(q)$. This form has its maximum value at $\omega = \beta = \omega_L(q)/2$.

 For $R_T$ we need a
simple model with a sign change at some $\omega_0$; we take
\begin{equation}
{\Delta R_T(q,\omega)\over3Aq^2} = 
{\Delta W_T(q)\over\omega_0^2}[ -\Theta(\omega_0 - \omega) + 
\Theta(\omega - \omega_0)\Theta(2\omega_0 - \omega)].
\end{equation}

\section{ Nuclear correlations and ($\vec p, \vec n$) data}

Using the model forms just described and the calculated values for 
$\Delta S_L$ and $\Delta W_L$, we find the
following estimates for the values of the $\Delta R_\alpha$
(normalized for comparison to experimental values). We find that
$\omega_L = 293$ MeV for the range of interest, 200 MeV/$c \leq q \leq$ 
400 MeV/$c$.
For the constant model (22) we find
\begin{equation}
\Delta R_L/3Aq^2 = 2.6 \times 10^{-4} {\rm MeV}^{-1}
\end{equation}
for all $\omega$,
while for the exponential model (23) we obtain an upper bound
\begin{equation}
\Delta R_L/3Aq^2 \leq 3.8 \times 10^{-4} {\rm MeV}^{-1}.
\end{equation}
For the model of $\Delta R_T$ in (24), assuming that $\omega_0 = \omega_L$,
the value for $\omega \leq \omega_L$ is given by
\begin{equation}
\Delta R_T/3Aq^2 = -4.5 \times 10^{-4} {\rm MeV}^{-1}.
\end{equation}

These estimates may be compared to 
the data from the ($\vec p, \vec n$) experiments in 
the range 240 MeV/$c \leq q \leq$ 380 MeV/$c$, which show 
peak values (see \cite{ab}, Fig. 2)
\begin{equation}
\max R_\alpha/3Aq^2 \simeq 1.2 - 1.7 \times 10^{-2} {\rm MeV}^{-1},
\end{equation}
dropping to $\simeq 0.5 \times 10^{-2} {\rm MeV}^{-1}$ towards the
ends of the range of energy losses. 
The quoted values are dependent on the scattering model used to extract the 
response functions: see \cite{aa4} for a different analysis of the same data.
But the magnitudes are reasonable; they are of the same order as
the value (17) of the quasifree peak for a Fermi
gas (nuclear matter) at these momenta.

Our estimates of the correlation contributions to the response functions
are then of the order of a few percent at the
peak values, and less than $10\%$ over the whole range of excitation energies,
$\omega \leq$ 150 MeV. 
These contributions are ${\it smaller}$ than the 
estimated uncertainties in the data quoted, including counting
($\leq 10\%$), experimental systematic ($6\% - 8\%$), and model uncertainties
in extraction ($20\%, 10\%$).

So these experiments are not presently accurate enough to measure the effects 
expected from nuclear correlations, which is unfortunate. 
What is curious about the ($\vec p, \vec n$) data is that the values of
$R_L$ (which are connected to the pion excess) are not hard to understand, but 
those for $R_T$ seem to be enhanced. This is not explained by any conventional
nuclear mechanism, and has suggested, e.g., evidence for rho enhancement in 
nuclei \cite{ai}.

\section{ Pion contributions to DIS}

The estimates of the distributions $\Delta R_L$ also have consequences for the 
analysis of deep inelastic scattering (DIS) or related processes on nuclear 
targets. The quantity of interest  is the pion momentum distribution 
for the target, $p_A(y)$, defined in (16). This quantity is integrated over 
$y$, weighted by the pion structure function, to give the pion contribution
to the DIS.
Integrating (16) over the three-momentum gives 
\begin{equation}
p_A(y) = {f^2 \over 16 \pi^2 m_\pi^2}y\int^\infty_{(My)^2}dq^2 
\int^{\omega_m}_0 d\omega {F^2(t) R_L(q,\omega)\over {(t + m^2)^2}},
\end{equation}
where  $y = (q_z - \omega)/M$.
The upper limit of the $\omega$-integral is given by 
\begin{equation}
\omega_m \equiv q - yM \geq q_z - yM.
\end{equation}
The main point is the strong effect of the upper limit 
$\omega_m$ on the contribution of the nuclear response function to the
$\omega$-integral.  
To illustrate the effect,
making a static approximation as in (14), consider
\begin{equation}
J(q,y) = {1\over 3Aq^2} \int^{\omega_m}_0 d\omega R_L(q,\omega).
\end{equation}
Clearly, $J(q,y) \leq S_L(q)$, the suppression increasing with $y$.
As an example, consider
$q =$ 400 MeV/$c$, for which the value of $S_L(q)$ = 1.11 is a maximum, as
is the pion excess, also.
One finds that the values of $J(q,y)$ are reduced below 1.0 for
$y \simeq 0.3$. 

The consequence is that the effect of excess pions which are associated with 
NN correlations may be sufficiently reduced by kinematic 
constraints to be inaccessible in DIS or dimuon experiments \cite{au}.

\section{ Conclusions and discussion}

So, what do we learn? We have seen that there is a theoretical connection 
based on the meson theory of NN interactions, between nuclear binding and
certain inelastic scattering experiments which have been regarded as
sensitive to the pion excess in nuclei. The connection depends on the
spin-isospin correlations in nuclei and the response functions $R_L$ and $R_T$
which characterize the scattering.

In the conventional theory of nuclear structure for which the ground states 
and excitation spectra are dominated by two-nucleon correlations, the
major changes in the response functions (compared to the quasifree response)
are expected to be small in magnitude, and spread over a large range of
$\omega$. As a consequence, we estimate that these changes in $R_L$ and $R_T$
are well within the uncertainties for the data
from the recent $(\vec p,\vec n)$ experiments, for $\omega \leq$ 150 MeV. 

In the analysis of DIS and dimuon production experiments, part of the 
high-$\omega$ tails for $R_L$ are cut off by kinematical constraints.
This can reduce -- even eliminate -- the effect of excess pions from the
structure functions. 

So, there is no apparent conflict between the conventional theory of the pion
excess and present experimental data, but there is no positive confirmation,
either. The response functions are still the key to testing the conventional
picture of nuclear correlations and mesons in nuclei, but the requirements
for experimental and theoretical accuracy are not easily met.


\begin{thebibliography}{99}
\bibitem{aa1} Daniel S. Koltun, Phys. Rev. C {\bf 57}, 1210 (1998).
\bibitem{ba} O. Benhar and V.R. Pandharipande, Revs. Mod. Phys. {\bf 65}, 817 
(1993).
\bibitem{ac} J.D. Sullivan, Phys. Rev. D {\bf 5}, 1732 (1972).
\bibitem{ad} V.R. Pandharipande, J. Carlson, S.C. Peiper, R.B. Wiringa,
and R. Schiavilla, Phys. Rev. C {\bf 49}, 789 (1994).
\bibitem{bf} K. Nishida and M. Ichimura, Phys. Rev. C {\bf 51}, 269 (1995).
\bibitem{aa4} A. De Pace, Phys. Rev. Lett. {\bf 75}, 29 (1995).
\bibitem{aa2} Adelchi Fabrocini, Phys. Lett. {\bf 322}, 171 (1994).
\bibitem{aa3} V.D. Efros, W. Leidemann, and G. Orlandini, Phys. Rev. Lett. 
{\bf 78}, 432 (1997).
\bibitem{ab} T.N. Taddeucci et al., Phys. Rev. Lett. {\bf 73}, 3516 (1994).
\bibitem{ai} G.E. Brown, M. Buballa, Z.B. Li and J. Wambach, Nucl. Phys. {\bf 
A593}, 295 (1995).
\bibitem{au} D.M. Alde et al., Phys. Rev. Lett. {\bf 64}, 2479 (1990).
\end{thebibliography}
\end{document}